\documentclass[a4paper,nofootinbib]{revtex4}
\usepackage{epsfig,amsmath,amssymb}
\def\Qz{{\bf Q}}
\def\qj{{\bf q}_j}

\def\tpm{$\pm$~}

\def\gs{\gamma_s}
\def\ee{$e^+e^-$~}
\def\pp{$pp$~}
\def\ppb{$p\bar p$~}
\def\AA{$AA$~}
\def\AuAu{Au-Au~}

\def\e{{\rm e}}

\def\d{{\rm d}}
\def\B{\boldmath}

\def\beq{\begin{equation}}
\def\eeq{\end{equation}}
\def\dndy{\langle \frac{\d n_j}{\d y} \rangle}
\def\muv{\boldsymbol{\mu}}
\def\CH{$\chi^2/{\rm dof}$}
\def\B{\boldmath}

\def\lsim{\raise0.3ex\hbox{$<$\kern-0.75em\raise-1.1ex\hbox{$\sim$}}}
\def\gsim{\raise0.3ex\hbox{$>$\kern-0.75em\raise-1.1ex\hbox{$\sim$}}}

\def\NP{{ Nucl.\ Phys.\ }}
\def\PL{{ Phys.\ Lett.\ }}
\def\PR{{ Phys.\ Rev.\ }}

\def\EP{{ Europ.\ Phys.\ J.\ C}}

\begin{document}

20.1.10

\title{A comparative analysis of statistical hadron production}

\author{F. Becattini}
\affiliation{Dipartimento di Fisica, Universit\`a di Firenze, and INFN Sezione di Firenze, Italy} 
\author{P. Castorina}
\affiliation{Dipartimento di Fisica, Universit{\`a} di Catania, and INFN Sezione di Catania, Italy}
\author{A. Milov}
\affiliation{Department of Particle Physics, Weizmann Institute of Science
Rehovot 76100, Israel}
\author{H. Satz}
\affiliation{Fakult\"at f\"ur Physik, Universit\"at Bielefeld, Germany}

\begin{abstract}
We perform a systematic comparison of the statistical model parametrization 
of hadron abundances observed in high energy \pp, \AA and \ee
collisions. The basic aim of the study is to test if the quality of 
the description depends on the nature of the collision process. In 
particular, we want to see if nuclear collisions, with multiple initial 
interactions, lead to ``more thermal'' average multiplicities than elementary 
\pp collisions or \ee annihilation. Such a comparison is meaningful 
only if it is based on data for the same or similar hadronic species 
and if the analyzed data has quantitatively similar errors. When 
these requirements are maintained, the quality of the statistical model 
description is found to be the same for the different initial collision 
configurations.
\end{abstract}

\maketitle

\section{Introduction}

One of the most striking observations in high energy multihadron production is that 
both species abundances and transverse momentum spectra (provided effects of
collective flow and gluon radiation are removed) follow the thermal pattern of an ideal 
hadron-resonance gas, with a universal temperature $T \simeq 160-170$ MeV 
\cite{intro}. This behavior was initially attributed to a temperature limit arising 
from an exponentially growing resonance spectrum \cite{Hagedorn,DRM} and is today 
generally taken as a reflection of the quark-hadron transition temperature in QCD 
\cite{Ca-Pa}. 

\medskip

Thermal multihadron production has been investigated experimentally in a variety of 
collision processes, from \ee annihilation and \pp-\ppb collisions to high energy 
nucleus-nucleus interactions. For sufficiently high energies, these studies all led 
to the same universal hadronic resonance gas temperature, even though 
there were other distinguishing features. In particular, it was observed that in 
elementary collisions, strangeness production suffered an overall suppression, quite 
likely due to the heavier mass of the strange quark. In nuclear interaction,
this suppression is less or perhaps completely absent\footnote{It has
recently been noted that whatever strangeness suppression remains in heavy ion
collisions can be accounted for by residual (``corona'') single nucleon-nucleon 
interactions \cite{corona}.}.

\medskip

The origin of the success of a thermal picture for such a variety of different 
collision configurations has been an enigma for a long time, extensively discussed 
in the literature \cite{diba}. In particular, given a large number of possible 
multi-hadron channels, why does nature always choose to maximize entropy at a universal 
temperature within a finite region? In heavy ion collisions, it is conceivable that 
this could have something to do with the relatively large volume, which makes the system 
confined for a long enough time to allow sufficient inelastic scattering to
reach equilibration. However, this cannot account for the observation of thermal 
behaviour in elementary collisions, where such a  mechanism cannot play any role;
also, a hadronic-rescattering thermalization is hardly reconciled with the observation
of a centrality-independent chemical freeze-out temperature in relativistic
heavy-ion collisions \cite{kestin}. These facts are a strong indication that 
the thermal behaviour is a feature of hadronization itself. 
One explanation proposed for a universal spontaneous thermal hadron emission is 
that it arises through quantum tunnelling (Unruh radiation) at the color event horizon 
\cite{CKS}; the Schwinger mechanism \cite{Schwinger} is a special case of such
Unruh radiation \cite{Brout}. Among other ideas put forward to explain the
mechanism of the apparent thermalization, it is worth mentioning quantum 
chaos and Berry's conjecture \cite{krz,intro}, still at a speculative stage. 

\medskip

In some recent studies \cite{bcms,pbm,csorgo}, the extent of agreement of data 
with a thermal description was discussed for different processes. In a 
``thermalization'' framework where inelastic collisions play a major
role, it is suggestive that elementary 
processes should not lead to as good an agreement to the same, though approximated, 
statistical model formula, as it is found for nuclear collisions, with a sort 
of hierarchy from \ee to \AA. Such a view would be supported if it was found that 
the agreement, as specified by e.g. \CH, is simply much better for \AA data than 
it is for that from \ee annihilation. This has in fact been claimed \cite{pbm}; 
however, if such comparisons are to be conclusive, several conditions have to be met.

\begin{itemize}
\item{First of all, any comparison should be based on essentially the
same set of species. Ideally, these should be narrow and hence 
long-lived resonance states, in order to avoid the difficulties
encountered in determining the rates of broad short-lived states,
concerning background separation, feed-down and branching ratios.
This requirement leaves in general some 10 to 15 states and so still 
allows a significant comparative analysis.}
\item{Next, if the comparison is to be based on the $\chi^2/{\rm dof}$
for the fits to be considered, the corresponding data sets should have 
similar experimental errors. If the estimated theoretical values are 
approximations, a more accurate data set may imply larger deviations 
between model and data in units of errors, so that a comparison of the 
\CH~ of two different sets makes sense only if their errors are comparable.}
\item{Finally, the basic idea of statistical hadronization can be 
implemented for specific process in different ways. Without decisive 
further information, a comparison of the quality of the fits remains the 
only tool to judge which scheme is closest to reality.}
\end{itemize}

We have here emphasized the use of \CH~as a tool to {\sl compare} different
experimental configurations as well as different model implementations.
The reason for this is, as we shall discuss in more detail in section III,
that the absolute value of \CH~has to be interpreted with much care. Since
the theoretical formulae employed in any statistical model fit are only
approximations of the full dynamics, deviations must appear, as has been
mentioned, once the measurements become sufficiently precise, and hence 
\CH~must then become large. This point was already made about 10 years
ago \cite{heinz}, when discussing the effect of local fluctuations
of thermodynamical parameters.

\medskip  

With these caveats in mind, we have selected three extensive data sets for our
comparative analysis. In Section II, following a short summary defining the 
details of the underlying statistical model framework, we compare recent 
hadroproduction data from \pp collisions, taken by the STAR experiment at
RHIC for $\sqrt s=200$ GeV \cite{starpp}, to the corresponding results from 
\AuAu collisions \cite{starau} at the same energy and measured by the same
experimental group. One very remarkable result of this analysis is, as we
shall see, that the fit to the \pp data is as good as it can possibly
be, with a \CH~$\simeq 1$; the corresponding \AuAu analysis leads to 
a less optimal fit. To elucidate the meaning of this result, we discuss
in Section III more generally the relevant features of comparing the
statistical hadronization model to data and testing the fits obtained.
In Section IV, we then consider \ee data from LEP at 91.25 GeV \cite{LEP}
and compare the fits for this to the \AuAu results. In Section V, we 
consider more generally tests of the different implementations of the 
statistical model; in this context, we also consider possible
origins of recent apparently contradictory conclusions \cite{pbm,bcms} 
on the thermal description of \ee data.

\section{A comparison between \B\pp and \B\AA collisions}

In this Section we perform a statistical analysis of hadroproduction 
data at $\sqrt s =200$ GeV at RHIC with \pp and \AuAu as initial 
collision configurations. The data in both cases are centre-of-mass 
midrapidity densities from the same experiment \cite{starpp,starau}.
For \pp collisions, there are 18 species of measured secondaries 
\cite{starpp}, including several short-lived strange meson and hyperon 
resonant states; for \AuAu interactions, we use a set of 12 rapidity 
densities of long-lived states at midrapidities \cite{starau}, already
studied in ref.~\cite{becamann}, with updated experimental errors. The 
observed abundances are listed in table~\ref{fitpp2}.

\begin{table}[!h]\begin{center}
\begin{tabular}{|c|c|c|c|c|c|}
\hline
     Particle                      & Measured $dN/dy$ (E)  & Relative error &  Model $dN/dy$ (M)  &  Residual  & (M - E)/E (\%)  \\                                    
\hline
 \multicolumn{6}{|c|}{pp collisions at $\sqrt s =$ 200 GeV} \\
\hline
$\pi^+$                            &  1.44  \tpm 0.11      &  0.076  &   1.403      &  -0.34   &  -2.62  \\ 
$\pi^-$                            &  1.42  \tpm 0.11      &  0.077  &   1.384      &  -0.33   &  -2.59  \\ 
$K^+$                              &  0.150 \tpm 0.013     &  0.087  &   0.1522     &   0.17   &  1.48   \\ 
$K^-$                              &  0.145 \tpm 0.013     &  0.090  &   0.1460     &   0.076  &  0.68   \\
$p$                                &  0.138 \tpm 0.012     &  0.087  &   0.1491     &   0.92   &  7.42   \\
$\bar p$                           &  0.113 \tpm 0.010     &  0.088  &   0.1120     &   0.66   &  5.56    \\
$\phi$                             &  0.0180 \tpm 0.0029   &  0.16   &   0.01130    &  -2.31   &  -59.3  \\
$\Lambda$                          & 0.0436 \tpm 0.0041    &  0.094  &   0.04348    &  -0.030  &  -0.28   \\
$\bar\Lambda$                      & 0.0398 \tpm 0.0038    &  0.095  &   0.03686    &  -0.77   & -7.96   \\
$\Xi^-$                            & 0.0026 \tpm 0.00092   &  0.35   &   0.003070   &   0.51   &   15.3  \\
$\bar\Xi^+$                        & 0.0029 \tpm 0.00104   &  0.36   &   0.002728   &  -0.17   &   -6.29  \\
$\Omega+\bar\Omega$                & 0.00034 \tpm 0.00019  &  0.56   &   0.0005712  &   1.22   &   40.5  \\
\hline
$K^0_S$                            & 0.134   \tpm 0.011    &  0.082    & 0.1467     &  1.15    &   8.64  \\
$\rho^0$                           & 0.259 \tpm 0.039      &  0.15     & 0.1861     & -1.87    &  -39.2  \\
$(K^{*0}+\bar K^{*0})/2$           & 0.0508 \tpm 0.0063    &  0.12     & 0.05151    &  0.11    &   1.38   \\
$\Sigma^{*+}+\Sigma^{*-}$          & 0.0107 \tpm 0.00146   &  0.14     & 0.01028    &  -0.29   &  -4.12    \\
$\bar \Sigma^{*+}+\bar \Sigma^{*-}$& 0.0089 \tpm 0.00126   &  0.14     & 0.008650   &  -0.20   &  -2.89   \\
$\Lambda(1520)+\bar\Lambda(1520)$  & 0.0069 \tpm 0.0011    &  0.16     & 0.005606   &  -1.18   &  -23.1   \\
\hline
 \multicolumn{6}{|c|}{Au-Au collisions at $\sqrt s_{NN}$ = 200 GeV} \\
\hline
$\pi^+$            & 322 \tpm   25       & 0.078   &   330.0       &    0.32   &     2.41   \\
$\pi^-$            & 327 \tpm   25       & 0.077   &   331.9       &    0.19   &     1.46   \\
$K^+$              & 51.3 \tpm    6.5    & 0.13    &    57.65      &    0.98   &    11.0  \\
$K^-$              & 49.5 \tpm    6.2    & 0.13    &    54.44      &    0.80   &     9.07   \\
$p$                & 34.7 \tpm    4.4    & 0.13    &    42.23      &    1.71   &    17.8  \\
$\bar p$           & 26.7 \tpm    3.4    & 0.13    &    31.24      &    1.34   &    14.5   \\
$\Lambda$          & 16.7  \tpm   1.12   & 0.067   &    14.44      &   -2.02   &   -15.7  \\
$\bar\Lambda$      & 12.7  \tpm   0.92   & 0.072   &    11.10      &   -1.74   &   -14.4  \\
$\phi$             &  7.95 \tpm   0.74   & 0.093   &     6.697     &   -1.69   &   -18.7  \\
$\Xi^-$            &  1.83  \tpm   0.206 & 0.092   &     2.024     &   -0.73   &    -7.20  \\
$\bar \Xi^+$       &  2.17  \tpm   0.20  & 0.11    &     1.676     &   -0.75   &    -9.16 \\
$\Omega+\bar\Omega$&  0.53  \tpm   0.057 & 0.11    &     0.6529    &    2.16   &    18.8 \\
\hline
\end{tabular}\end{center}
\caption{Measured and fitted mid-rapidity densities in \pp and \AuAu
collisions at 200 GeV; data from STAR experiment. For \pp collisions, none of the quoted 
experimental numbers are corrected for weak decay feed-down \cite{starpp}, while 
for Au-Au collisions all multiplicities are feed-down corrected, except protons and 
antiprotons \cite{starau}. Our model calculations were carried out accordingly.}
\label{fitpp2}
\end{table}

\medskip

For elementary collisions, the abundance $\langle n_j \rangle$ of hadron
species $j$ is in the statistical hadronization model given by
(for a detailed description, see ref.~\cite{bcms})
\beq\label{form}
 \langle n_j \rangle ^{\rm primary} = \frac{V T (2J_j+1)}{2\pi^2} 
 \sum_{n=1}^\infty \gs^{N_s n}(\mp 1)^{n+1}\;\frac{m_j^2}{n}\;
 {\rm K}_2\left(\frac{n m_j}{T}\right)\, \frac{Z(\Qz-n\qj)}{Z(\Qz)},
\eeq
where the temperature $T$, the strangeness suppression $\gs$ and the
normalization volume $V$ are taken as free parameters; $\Qz = (Q,B,S,\ldots)$ 
is the array of conserved charges and $\qj$ the corresponding array for the 
$j$th hadron species. The ``chemical'' factors $Z(\Qz-n\qj)/Z(\Qz)$ are ratios 
of partition functions and replace the more familiar fugacities when 
the exact conservation of the initial charges is to be taken into account, 
a typical feature of small systems also known as canonical suppression. 
To the primary production for each species $j$ one then adds the decay 
products of heavier states, using the experimentally known branching ratios 
\beq\label{deca}
  \langle n_j \rangle = \langle n_j \rangle^{\rm primary} +
  \sum \langle n_k \rangle BR(k \to j).
\eeq
Formulae (\ref{form}) and (\ref{deca}) apply in principle to full phase 
space multiplicities, since possible charge-momentum correlations are integrated 
out. In order to apply it nevertheless in a comparison of midrapidity data from 
\pp and \AuAu collisions at the same energy, we assume that particle ratios 
at midrapidity are essentially the same as particle ratios in full phase space. 
While such an assumption is certainly not tenable at low collision energy, it might 
improve its validity in high energy collisions with large rapidity coverage. We 
thus assume that the primary rapidity density of each species in \pp collisions 
is given by
\beq\label{form2}
 \dndy^{\rm primary}_{y=0} = 
 \frac{A V T (2J_j+1)}{2\pi^2} 
 \sum_{n=1}^\infty \gs^{N_s n}(\mp 1)^{n+1}\;\frac{m_j^2}{n}\;
 {\rm K}_2\left(\frac{n m_j}{T}\right)\, \frac{Z(\Qz-n\qj)}{Z(\Qz)},
\eeq
where $A$ is a common normalization factor taking into account the ratio of production 
in the mid-rapidity interval to the overall rate. Note that $A$ cannot be absorbed 
into the overall volume $V$, since $V$ still appears as a key parameter in the 
chemical factors $Z(\Qz-n\qj)/Z(\Qz)$. Also, it is assumed that subsequent 
decays do not alter the midrapidity densities, i.e. that also formula (\ref{deca}) 
holds. 

\medskip

In this form, the statistical model for \pp collisions leads to a
four-parameter fit: besides $T$, $\gs$ and $V$, there is the rapidity-cut
related normalization factor $A$. For heavy ion collisions, we have
exactly the same number of parameters; because of the large volume, formula
(\ref{form2}) here becomes
\beq\label{formh}
 \dndy^{\rm primary}_{y=0} = 
 \frac{A V T (2J_j+1)}{2\pi^2} 
 \sum_{n=1}^\infty \gs^{N_s n}(\mp 1)^{n+1}\;\frac{m_j^2}{n}\;
 {\rm K}_2\left(\frac{n m_j}{T}\right)\, \e^{n \muv \cdot \qj/T},
\eeq
with the chemical factors now replaced by the fugacities. In this case, the factor 
$A$ can be absorbed in $V$ and the free parameters are $T$, $\gs$, $\mu_B$
and $AV$.

\begin{table}[!h]\begin{center}
\begin{tabular}{|c|c|c|}
\hline
                   & pp $\sqrt s = 200$ GeV      & Au-Au $\sqrt s_{NN}= 200$ GeV  \\                                   
\hline
 \multicolumn{3}{|c|}{\bf Overall fit} \\
\hline
 T(MeV)            &  170.1\tpm 3.5              &  168.5 \tpm 4.0             \\                                   
 Normalization     &  0.027\tpm 0.011            &  13.6 \tpm 0.58             \\
 $VT^3$            &  135 \tpm 60                &                             \\                    
 $\gamma_S$        &  0.569\tpm 0.031            &  0.932 \tpm 0.040           \\
 $\mu_B/T$         &                             &  0.173 \tpm 0.052          \\                                     
 $\chi^2/dof$      &  15.6/14                    &  22.2/8                    \\                                     
\hline
 \multicolumn{3}{|c|}{\bf Fit with standard sample} \\
\hline
 T(MeV)            &  169.8\tpm 4.2              &  168.5 \tpm 4.0           \\                                    
 Normalization     &  0.028\tpm 0.012            &  13.6 \tpm 0.58           \\ 
 $VT^3$            &  131 \tpm 60                &                            \\
 $\gamma_S$        &  0.600\tpm 0.033            &  0.932 \tpm 0.040         \\                                     
 $\mu_B/T$         &                             &  0.173 \tpm 0.052         \\                                     
 $\chi^2/dof$      &  15.0/8                     &  22.2/8                  \\  
\hline
\multicolumn{3}{|c|}{\bf Fit with standard sample and same relative errors} \\
\hline
 T(MeV)            &  170.9\tpm 5.2              &  169.6 \tpm 4.8            \\                                    
 Normalization     &  0.031\tpm 0.015            &  12.6 \tpm 0.73            \\
 $VT^3$            &  118 \tpm 63                &                            \\
 $\gamma_S$        &  0.595\tpm 0.038            &  0.999 \tpm 0.069          \\                                     
 $\mu_B/T$         &                             &  0.163 \tpm 0.065         \\                              
 $\chi^2/dof$      &  13.4/8                     &  9.6/8                    \\                              
\hline
\end{tabular}\end{center}
\caption{Comparison between the statistical model fits in \pp and \AuAu collisions
at $\sqrt s = 200$ GeV. The parameter referred to as ``Normalization" is $A$ in
eq.~(\ref{form2}) for \pp collisions and $AV T^3 \exp(-0.7 GeV/T)$ in eq.~(\ref{formh}) 
for Au-Au collisions.}
\label{fitpp}
\end{table}
\medskip

\begin{figure}[h]
\centerline{\psfig{file=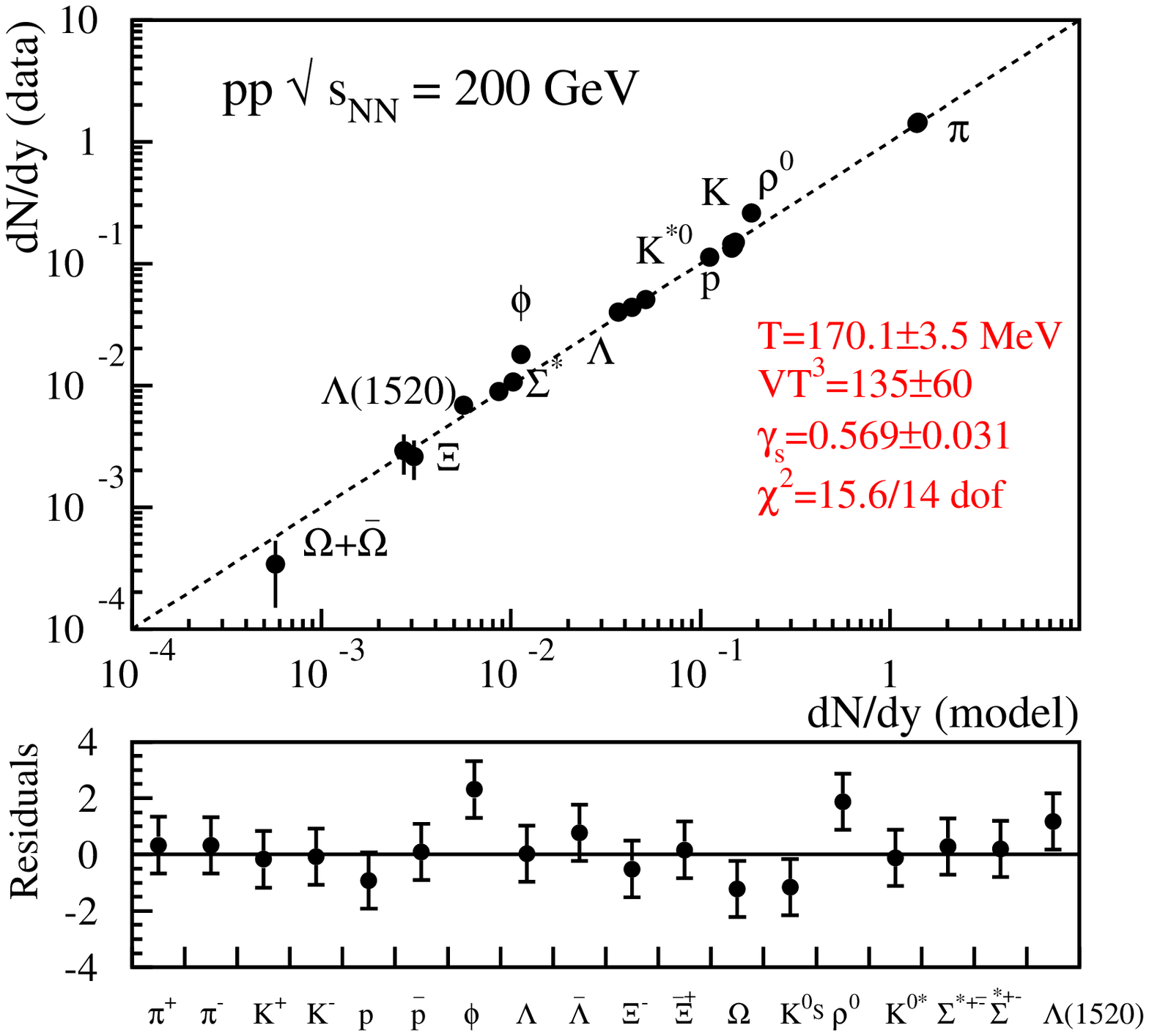,width=9cm}}
\caption{Above: fitted vs measured midrapidity densities in \pp collisions
at $\sqrt s= 200$~GeV. Below: residual distributions.} 
\label{fitpp1}
\end{figure}

The results of the fits to the midrapidity densities are shown in table~\ref{fitpp2} 
for both \pp and \AuAu and fig.~\ref{fitpp1}; the fit parameter values and the 
corresponding \CH~ are listed in table~\ref{fitpp}. The first block of this table, 
labelled ``full \pp fit", gives the parameters for the full analysis of 
all \pp species. In the middle block of table~\ref{fitpp},
we then compare the $pp$ and \AuAu fits using the same data sample.  
 
\medskip

The most striking feature seems to us the high quality of the full \pp fit.
The resulting \CH~$\simeq 1$ is the best value to be hoped for; it is as good
as any thermal fit ever made for any high energy collision 
configuration.\footnote{A recent analysis \cite{kcor} of $p\!-\!p$ data at a 
considerably lower energy, $\sqrt s=17$ GeV, also finds very good agreement
with a different implementation of the statistical model.} 

\medskip

Next, it is worth noting that the extracted temperature values are 
almost identical for \pp and \AuAu collisions; the $\gs$ value
for the \pp data agree with previous \pp analyses \cite{becapp, becapt}.
The error on $VT^3$ here is large because this parameter is essentially 
determined by the chemical factors in eq.~(\ref{form2}). 

\medskip

In spite of the extremely similar values obtained for the hadronization 
temperature,  we thus find that the \CH~of the \pp fits is a factor 
two better than that for the \AuAu analysis. A naive conclusion of 
this comparison could then be 
that the statistical model leads to a better agreement with \pp collisions 
than for $Au-Au$. Can we conclude that $pp$ collisions provide a "more
thermal" configuration than heavy ion collisions? Before answering this 
question, a general discussion about the statistical model formulae and 
the meaning of statistical tests is now appropriate.

\section{Theoretical models and \B$\chi^2$ tests}

Here we want to discuss more in detail two points that have been mentioned 
in the Introduction, namely the meaning of a \CH~fit given only an approximate 
theoretical description, and how one can compare two fits to such an incomplete 
input.

\medskip

In general, the test of a physical model is usually based on a fit of data and a 
statistical test of this fit. Thus a $\chi^2$ test tells us how 
likely it would be to obtain a value larger than the minimum $\chi^2$ 
of our fit, provided that the hypothesis, i. e., the model, is correct. 
Most often, however, the theoretical formulae that we take as hypotheses 
are only an approximate representation of the underlying model or theory. 
In other words, they can be expected to reproduce the data only up to some 
reasonably small deviation. There are many instances of this situation;
a simple example is the lowest-order perturbative expansion of a differential 
cross section in high-energy collisions. If measurements are more accurate 
than the estimated deviation, the $\chi^2$ statistical test will obviously 
fail, indicating that corrections to the lowest-order
theoretical formula are necessary. Sometimes the theoretical description 
of the process is fully under control and corrections are relatively easy 
to calculate (as for electroweak processes in \ee collisions), sometimes 
they are not. This latter is the case in this work, where we test the 
statistical hadronization model.

\medskip

The basic premise of the statistical model is that high energy collisions lead
to the formation of multiple clusters, emitted sequentially in rapidity and decaying
into hadrons according to their relative phase space weights. The formulae
(\ref{form}) and (\ref{formh}) are specific implementations of this idea,
based on further additional assumptions besides the basic postulate 
of the model. In elementary collisions, it is assumed that the probability 
of distributing the conserved charges among the actually produced clusters 
has a special form: for instance, a statistical distribution of charges among
the clusters, leading to the equivalence with one global cluster (this is
assumed in our work here). But other charge distribution schemes are obviously
conceivable, see ref.\ \cite{beca}, and we shall return to this aspect in
Section V. It is therefore clear that if experimental multiplicities were known 
to a very good accuracy, discrepancies with the predictions or fitted values 
of formula (\ref{form}) could show up, even if the basic idea of purely 
statistical decays of clusters/fireballs remains true. One would then have to 
correct (\ref{form}) for effects of a non-statistical distribution of 
charges among the clusters, etc. Unfortunately, the definition of such
corrections requires a more complete picture of the production process than we 
presently have. Of course, this does not mean that approximate models cannot
be disproved. It only means that, as long as corrections to the leading-order 
formula are not available, we have to be content with a statement like "the 
statistical hadronization model in its simplest approximation reproduces the
data up to 10\%", and we should take deviations from formula (\ref{form}) with 
great care.

\medskip

Above and beyond these model-dependent details, we should also stress
that any analytical multiplicity formulae, such as (\ref{form}) and
(\ref{formh}) or variations thereof, provide by construction only an 
approximate description. The use of these formulae will necessarily lead 
to deviations for sufficiently precise measurements. To illustrate this 
situation, which is ubiquitous in physics, we recall a familiar example. 
The spectrum of the sunlight measured at the top of the atmosphere is usually 
fitted to a black body formula, yielding a temperature of about
5780$^{\circ}$K. The formula is accurate to about 10\%, but the 
fit quality is extremely bad, with a huge $\chi^2$ (see
e.g. ref.~\cite{sunspec}). 
The reason for such discrepancy is that on a microscopic level, stars are not
perfect radiators; different effects, such as the absorption of light by atoms 
and/or ions blocking part the outward radiation path, surface non-uniformities, 
etc., all cause deviations from the simple black-body spectrum, and these are 
revealed when the accuracy of the photometric measurement is better than about 
10\%. The effects responsible for these deviations are difficult to embody in an 
analytical formula and can only be studied numerically. Still, the 
failure of the lowest-order Planck formula fit in passing a rigorous statistical 
$\chi^2$ test has not led anyone to the conclusion that the surface of the sun 
is not a thermal system. 

\medskip

In addition to such theoretical caveats, one must also bear in mind
the role of experimental complications. In fact, most of the measurements
which are used to perform a fit include hidden correlations which are
difficult to assess (e.g., multiplicities of particles measured with the 
same detector). Therefore the usual assumption of independent errors 
entering in the $\chi^2$, which we have retained here throughout, is only an 
approximation, and the actual absolute value of the best \CH~could well
be different.

\medskip

Hence, for theoretical as well as for experimental reasons, the use of
absolute \CH~values to judge the quality of statistical model descriptions, 
as suggested in ref.\ \cite{csorgo}, appears to us as not really permissible.

\medskip

But even with these caveats in mind, a comparative assessment of fit results in
different collision configurations is still possible. To be specific, if the 
fit quality to a given formula was better in \AA collisions than in \ee or 
\pp interactions under the same conditions, this could of course mean that the 
deviations observed in \ee and \pp stem from a genuine failure of the model
for this case, rather than from the approximate nature of the employed formula. 
However, for such a comparison to make any sense, it must fulfill the
essential prerequisite noted in the Introduction. Thus, if the comparison is 
to be based on a $\chi^2$ test for the fits to be considered, the
corresponding data sets must have comparable average experimental errors. 
It is quite clear that if data sets with largely different accuracy are used, 
a comparison of \CH~values could be misleading. To illustrate: if the
data set $A$ (example: hadronic multiplicities in \pp) with an average
accuracy of 10\% yields a fit with \CH~= 1 and the set $B$ (example: 
hadronic rapidity densities in heavy ion collisions) with an average accuracy
of $1\%$ yields a fit with \CH~= 3, a blind comparison of $\chi^2$'s 
would lead us to the erroneous conclusion that the model can be used to describe 
$A$, but not $B$. 

\medskip

We now return to our comparison of Sect.~2 and the hypothetical conclusion
that the statistical model works better in \pp collisions than in \AuAu 
collisions. A comparison of $\chi^2$ values requires that measurements have 
the same experimental accuracy, but this is not the case here: as can be 
deduced
from table~\ref{fitpp1}, the relative experimental errors for the same species
sets differ, with an average value of about 18\% for the \pp data but only 
10\% for \AuAu.

\medskip

A possible way of comparing the quality of the two fits would be to look at
the average relative deviation between theoretical and experimental values. 
For each particle species this is shown in the last column of table~\ref{fitpp1}. 
Indeed, the average deviation of the 12 common species are very similar: 12.5\% 
in pp and 11.7\% in Au-Au. This result is a strong indication that the
statistical model indeed yields the same quality of agreement in the examined cases. 
To illustrate this more precisely, we have made new fits with the experimental
errors rescaled such that the relative errors of the measurements are exactly 
equal and determined by the largest error for each species. For instance, the 
$K^+$ multiplicity in \pp collisions is now assigned an error equal to its 
relative error in heavy ion collisions times its multiplicity in \pp
collisions, that is $0.13 \times 0.150 = 0.0195$, giving it the same relative error 
as the corresponding measurement in \AuAu collisions. Conversely, for the 
$\Omega + \bar \Omega$ in \AuAu collisions, the measurement is far more accurate 
than that in pp collisions; hence here a larger error is assigned to the heavy 
ion value in the same manner. This procedure artificially enhances the experimental 
errors in both samples and can thus be used only for illustrative purposes, 
neither to extract the best estimate of the model parameters, nor to make a
proper statistical test. 

\medskip

The results of these fits are shown in table~\ref{fitpp2} (third block), and they 
show that the \CH~in \AuAu is now slightly better than in \pp collisions at the 
same energy. This reinforces our first assessment that the simple statistical model
formulae (\ref{form2}) and (\ref{formh}) agree with the data up to $\approx
10\%$ both in \pp and \AuAu collisions and that none of the examined
systems can be claimed to be ``more thermal" in this respect.

\section{A comparison between \B$e^+e^-$ and \B\AA collisions}

Here we focus our attention on the largest and the most accurate \ee sample,
data from LEP at $\sqrt s = 91.25$ GeV; this we compare again to the heavy ion
sample from RHIC at $\sqrt s_{NN}=200$ GeV, and we again choose as common as
possible a set of long-lived particles. Since in \ee collisions, the
multiplicities of particle and antiparticle are 
obviously equal, we now find only 7 common species
($\pi^\pm$, $K^\pm$, $\Lambda$, $p$, $\phi$, $\Xi^\pm$, $\Omega$). Therefore 
we have retained all 12 RHIC measurements ($\pi^\pm$, $K^\pm$, $\Lambda$, $\bar\Lambda$,
$p$, $\bar p$, $\phi$, $\Xi^\pm$, $\Omega+\bar\Omega$) and added to the \ee data sample 
$\pi^0$, $K^0_S$, and the three long-lived $\Sigma$ states.

\medskip

The experimental values used for this comparison are a weighted average
of the full phase space multiplicities measured by the four LEP experiments;
these were also used in our previous analysis\cite{bcms}\footnote{We note
that these weighted averages differ slightly from those compiled by the 
Particle Data Group \cite{pdg}. In particular, in most cases our errors 
are slightly smaller than those quoted in ref.~\cite{pdg} which makes the 
comparison even more conservative with regard to our final conclusions.}. 
The resulting fit is to be compared to the \AuAu mid-rapidity densities 
measured by STAR \cite{starau} and already used above. For comparison 
purposes, they are shown again in table~\ref{fitee}, together with the
corresponding \ee fit values.

\begin{table}[!h]\begin{center}
\begin{tabular}{|c|c|c|c|c|c|}
\hline
                   & Measured $dN/dy$ (E)  & Relative error &  Model $dN/dy$ (M)  & Residual    & (M - E)/E (\%)  \\                                   
\hline
 \multicolumn{6}{|c|}{\ee collisions at 91.25 GeV} \\
\hline
$\pi^0$            &  9.61  \tpm  0.29     & 0.030    &  9.865         &  0.89     &   2.6      \\                                  
$\pi^+$            &  8.50  \tpm  0.10     & 0.012    &  8.460         & -0.37     &  -0.44      \\                                 
$K^+$              &  1.127 \tpm 0.026     & 0.023    &  1.080         & -1.79     &  -4.3      \\                                  
$K_S^0$            &  1.0376 \tpm  0.0096  & 0.0093   &  1.040         &  0.25     &   0.24      \\                                 
$p$                &  0.519 \tpm   0.018   & 0.035    &  0.5727        &  2.98     &   9.4     \\                               
$\phi$             &  0.0977 \tpm  0.0058  & 0.059    &  0.1179        &  3.47     &   17.1     \\                              
$\Lambda$          &  0.1943 \tpm  0.0038  & 0.020    &  0.1867        &  -1.98    &   -4.0     \\                              
$\Sigma^{+}$       &  0.0535 \tpm  0.0052  & 0.097    &  0.04331       &  -1.96    &   -23.6    \\                              
$\Sigma^{0}$       &  0.0389 \tpm  0.0041  & 0.11     &  0.04392       &  1.22     &    11.4    \\                              
$\Sigma^{-}$       &  0.0410 \tpm  0.0037  & 0.090    &  0.03949       &  -0.40    &    -3.7    \\                              
$\Xi^-$            &  0.01319 \tpm 0.00050 & 0.038    &  0.01285       &  -0.65    &    -2.7    \\                              
$\Omega$           &  0.00062 \tpm 0.00010 & 0.16     &  0.0009264     &  2.55     &    33.0    \\
\hline
 \multicolumn{6}{|c|}{Au-Au collisions at 200 GeV} \\
\hline
$\pi^+$            & 322 \tpm   25       & 0.078   &   330.0       &    0.32   &     2.41   \\
$\pi^-$            & 327 \tpm   25       & 0.077   &   331.9       &    0.19   &     1.46   \\
$K^+$              & 51.3 \tpm    6.5    & 0.13    &    57.65      &    0.98   &    11.0  \\
$K^-$              & 49.5 \tpm    6.2    & 0.13    &    54.44      &    0.80   &     9.07   \\
$p$                & 34.7 \tpm    4.4    & 0.13    &    42.23      &    1.71   &    17.8  \\
$\bar p$           & 26.7 \tpm    3.4    & 0.13    &    31.24      &    1.34   &    14.5   \\
$\Lambda$          & 16.7  \tpm   1.12   & 0.067   &    14.44      &   -2.02   &   -15.7  \\
$\bar\Lambda$      & 12.7  \tpm   0.92   & 0.072   &    11.10      &   -1.74   &   -14.4  \\
$\phi$             &  7.95 \tpm   0.74   & 0.093   &     6.697     &   -1.69   &   -18.7  \\
$\Xi^-$            &  1.83  \tpm   0.206 & 0.092   &     2.024     &   -0.73   &    -7.20  \\
$\bar \Xi^+$       &  2.17  \tpm   0.20  & 0.11    &     1.676     &   -0.75   &    -9.16 \\
$\Omega+\bar\Omega$&  0.53  \tpm   0.057 & 0.11    &     0.6529    &    2.16   &    18.8 \\
\hline
\end{tabular}\end{center}
\caption{Measured and fitted $4\pi$ multiplicities in \ee collisions at $\sqrt s = 91.25$ GeV
and mid-rapidity abundances in \AuAu collisions at 200 GeV. For \ee collisions, all quoted 
experimental numbers (for references see ref.~\cite{bcms}) include weak decays feed-down, 
while for Au-Au collisions all multiplicities are feed-down corrected but protons and 
antiprotons \cite{starau}. Our model calculations were carried out accordingly.}
\label{fitee}
\end{table}

\medskip

It should be noted that in the \AuAu collisions at $\sqrt s_{NN} = 200$ GeV the 
experimental relative error is, in most cases, larger than for \ee collisions at LEP 
energy; the average error in the \ee data is 5.7\%, to be compared to 10\% for 
the heavy ion data. This suggests that the fit will result in a larger
\CH~value for \ee than for \AuAu, and we see in table~\ref{fitee2}
that this is indeed the case\footnote{The \CH~value obtained here is slightly
larger than that found in \cite{bcms}; this is due to the fact, unlike in 
ref.~\cite{bcms}, we have not included in the $\chi^2$ the contribution to errors 
owing to the uncertainties on masses, widths and branching ratios of resonances.
This choice is motivated by the need of making the comparison between different 
collision systems as clean as possible.}. Therefore, as emphasized above, a direct
comparison of the \CH~values to assess the relative quality of the
two fits would again be misleading. 

\begin{table}[!h]\begin{center}
\begin{tabular}{|c|c|c|}
\hline
                   & \ee $\sqrt s = 91.25$ GeV   & Au-Au $\sqrt s_{NN}= 200$ GeV  \\                                   
\hline
 \multicolumn{3}{|c|}{\bf Fit with the standard samples} \\
\hline
 T(MeV)            &  164.7\tpm 0.9 (1.9)        &   168.5 \tpm 4.0              \\                             
 Normalization     &  23.2\tpm 0.57 (1.2)        &   13.6 \tpm 0.58                \\
 $\gamma_S$        &  0.656\tpm 0.0096 (0.021)   &   0.932 \tpm 0.040              \\
 $\mu_B/T$         &                             &   0.173 \tpm 0.052             \\
 $\chi^2/dof$      &  41.5/9                     &   22.2/8                     \\
\hline
 \multicolumn{3}{|c|}{\bf Fit with the standard samples and same relative errors} \\
\hline
 T(MeV)            &  168.8\tpm 5.2              &  167.8 \tpm 4.1            \\                                    
 Normalization     &  21.3\tpm 3.4               &  13.15 \tpm 0.61            \\
 $\gamma_S$        &  0.599\tpm 0.029            &  0.968 \tpm 0.044          \\                                     
 $\mu_B/T$         &                             &  0.200 \tpm 0.057          \\                                     
 $\chi^2/dof$      &  11.0/9                     &     16.8/8                  \\                                    
\hline
\end{tabular}\end{center}
\caption{Comparison between fit results in \ee collisions at LEP and Au-Au 
collisions at RHIC for a fit to a sample of 12 long-lived hadronic species.
The parameter referred to as ``Normalization" is $VT^3$ for \ee collisions
and $AV T^3 \exp(-0.7 GeV/T)$ (see eq.~(\ref{formh})) for \AuAu collisions.
The errors within brackets are the fit errors rescaled by $\sqrt{\chi^2/dof}$
(see \cite{bcms}).}
\label{fitee2}
\end{table}

\medskip

As in the comparison between \pp and \AuAu~data, the average relative deviations 
between theoretical and experimental values are close: 9.4\% for \ee annihilation 
at 91.25 GeV and 11.7\% for Au-Au collisions at 200 GeV. 
This suggests again that the lowest order statistical model formulation yields 
the same quality of agreement in the two examined cases. To reinforce this, for
illustrative purpose, we have also here made new fits with rescaled experimental 
errors, in order to make the relative errors of measurements equal in the two samples 
for each species, in much the same way as for the comparison between \pp and \AuAu
collisions described in Sect.~III.
For the unmatched particles, we have taken an error correspondance $\pi^0 \to \pi^-$,
$K^0_S \to K^-$, $\Sigma^+ \to \bar p$, $\Sigma^0 \to \bar \Lambda$, $\Sigma^-
\to \bar \Xi^-$, where the first particle belongs to the \ee sample and the second 
to heavy ion sample. In fact, the only particles which have a larger error in
\ee than in \AuAu collisions are the $\Omega$ and $\Sigma^0$ , whose
correspondent in Au-Au was taken as $\bar\Lambda$.

\medskip

The resulting fit parameters are included in table~\ref{fitee2}. It is seen
that the final \CH~is now lower in \ee collisions than in \AuAu collisions,
although no strong statement can be made on this basis, as emphasized at the
end of previos section. This exercise only demonstrates that, under equal error 
conditions, the agreement of the data with the statistical model predictions in 
the form of eq.~(\ref{form}) is approximately the same in \ee and \AuAu collisions. 
Finally, we note that this result is found to be consistently independent of finer 
details, such as different particle species matching, number of particles, etc.

\section{Comparing different statistical descriptions of \ee collisions}

An interesting question of quite general interest is to what extent a $\chi^2$
test can be used to check if a specific {\em model} input is correct or not.
In other words, we now want to 
compare the \CH~values obtained by fitting a specific data 
set to different theoretical schemes, rather than a given theoretical scheme to 
different data sets. This is again a well-posed question because it involves only 
a comparative assessment. More specifically, we can fit the data to different 
implementations of the statistical model, some of which are likely to be or are 
certainly incorrect. If the model indeed reflects the right physics, the \CH~of 
a fit to the data should be consistently larger for ``incorrect'' versions than 
for the most realistic one. 

\medskip 

The \ee data from LEP at 91.25 GeV contains enough species and is sufficiently 
precise to address this issue. Our test set will consists of 15 light-flavored, 
long-lived particles, having widths less than 10 MeV: $\pi^0$, $\pi^\pm$, $\eta$, 
$\eta'$, $K^+$, $K^0_s$, $\phi$, $p$, $\Lambda$, $\Sigma^+$, $\Sigma^0$, $\Sigma^-$, 
$\Xi^-$, and $\Omega^-$. We now fit the measured abundances of these species to different 
implementations of the statistical model, using as basis eq.\ (\ref{form}).
\begin{itemize}
\item{}We fit the abundances to the primary production form only, neglecting
all resonance decay contributions.
\item{}We fit the abundances to the primary production form only, neglecting
all decay contributions from resonances of width greater than 10 MeV. In this
case, strong decays are neglected, but the feed-down of weakly decaying
heavy flavor states is included.
\item{}We fit the abundances correctly taking into account all resonance
decays, but we replace exact quantum number conservation (canonical
suppression) by a grand canonical formulation. This means replacing 
the chemical factors $Z({\bf Q}-{\bf q})/Z({\bf Q})$ in formula (\ref{form})
by the fugacities of (\ref{formh}).  
\end{itemize}
The results are summarized in table~\ref{alternative}; we have here denoted
the results from our global cluster formulation with ``correct'' inclusion
of all effects as ``full global implementation''. It is evident that the
model indeed provides us with a \CH~hierarchy: the cruder the implementation,
the larger the resulting \CH.

\begin{table}[!h]
\begin{center}
\begin{tabular}{|c|c|c|c|}
\hline
 Fit condition   & T (MeV) & $\gamma_S$ & $\chi^2$/dof   \\
\hline
 No resonance feed-down            & 158.7 &  0.494 &  1110/12  \\
 No strong resonance feed-down     & 161.1 &  0.537 &  391/12   \\
 No canonical suppression          & 144.9 &  0.690 &  131/12   \\
 Full global implementation        & 164.8 &  0.654 &  44/12    \\
\hline
\end{tabular}
\caption{Comparison of parameters for fits of 15 long-lived particles 
in \ee collisions at $\sqrt s = 91.2$ GeV, using different model inputs.}
\label{alternative}
\end{center}
\end{table}

\medskip
 
At this point, a specific feature of \ee annihilation is worth being emphasized. The 
annihilation process leads to the production of a pair of quarks, and about 40\% 
of \ee annihilations produce a pair of primary $c$ or $b$ quarks, as predicted by
the standard model. These then hadronize into heavy flavored hadrons, which
in turn predominantly decay into strange hadrons. The secondary production of 
light-flavored particles from heavy flavoured ones is a sizeable fraction of 
the overall particle production. It is thus crucial that the model correctly
includes the (non-statistical) fractions of the different primary quark pairs,
and fits neglecting these (e.g., in ref.\ \cite{pbm} all except the data set at 
91.2 GeV) cannot be considered as realistic.

\medskip 

\begin{figure}[h]
\centerline{\psfig{file=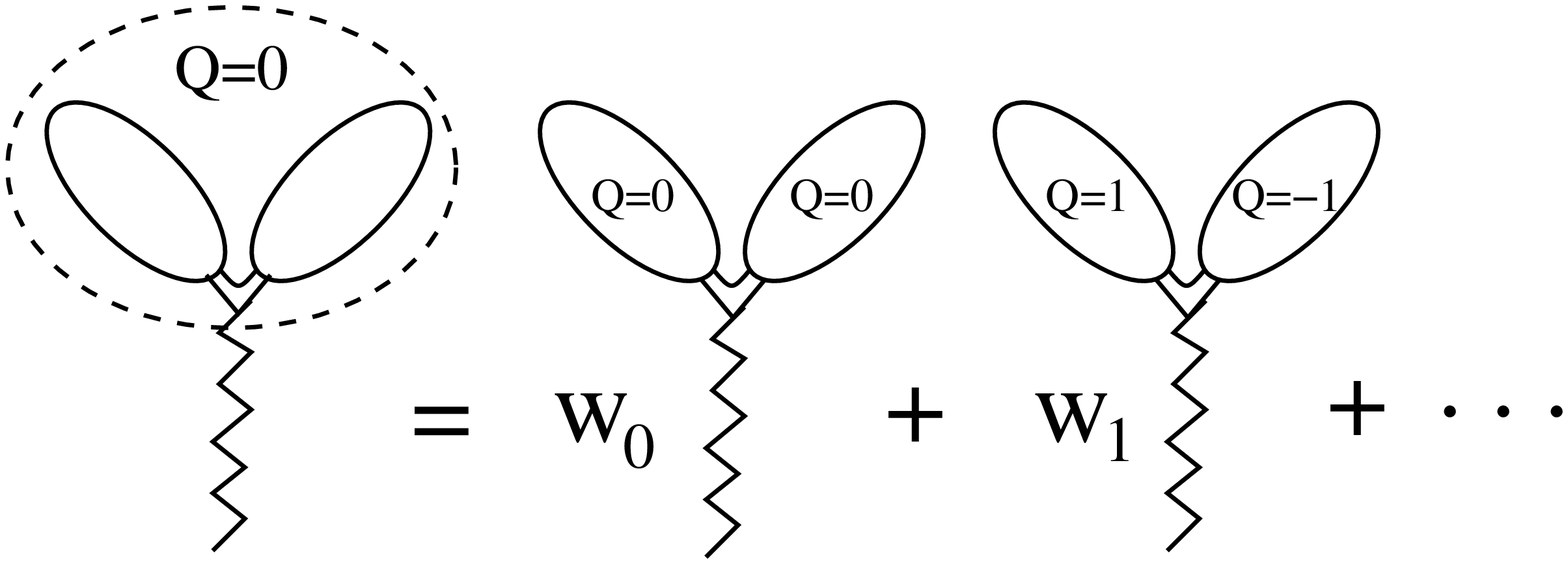,width=9cm}}
\caption{Schematic distribution of conserved quantum numbers for 
two-jet hadron production} 
\label{jets}
\end{figure}

Besides this aspect, however, there are others which distinguish different 
implementations of the statistical model concept. As already
indicated in Section III, the distribution pattern of the conserved
charges (baryon number, electric charge, strangeness, charm, bottom) is
an issue to be decided when formulating a specific model. If a given
overall quantum number (for illustration, consider the electric charge $Q$)
in two-jet production is zero, the general production pattern will lead to
a superposition of two produced jets each having zero charge, a pair 
with $Q=\pm 1$, one with $Q=\pm 2$, etc., as schematically illustrated in
Fig.\ \ref{jets}. To specify the model, we have to fix the weights $w_i$ 
of this superposition, and two particular schemes have been introduced 
for this \cite{beca}. 

\medskip

\begin{itemize}
\item{}The simplest version is to enforce exact conservation 
of all discrete quantum numbers separately for each of the two jets formed in the 
annihilation, allowing no quantum number exchange between the jets;
i.e., one puts $w_0=1$, $w_i=0 ~\forall~ i~\geq~ 1$. This is denoted as 
uncorrelated jet scheme.
\item{}The model we have used here allows clusters with
statistically distributed discrete quantum numbers, imposing exact
overall conservation laws; i.e, the $w_i$ are distributed over $i$
according to the available multicluster phase space for an overall 
$Q=0$. This is denoted as global cluster scheme. 
\end{itemize}
There is a clear exception, however, to a transfer of quark
quantum numbers: the heavy $c$ or $b$ quarks cannot be exchanged,
since heavy quark production at hadronization is severely suppressed. 
In fact, it is an experimentally well established fact that heavy quark production 
originates entirely from the primary \ee annihilation or, as an almost negligible higher 
order perturbative correction, from a hard gluon. It is also a well established fact 
that primary heavy quarks show up in open-heavy flavored hadrons without 
reannihilating. 

\medskip

There seem to be no general physics grounds on which to exclude one or the
other of these scenarios. However, color neutralization of the clusters 
created in \ee collisions requires, before hadron formation, the exchange of one 
or more quark-antiquark pairs between different clusters, and this in turn implies 
the appearance of non-vanishing integer additive charges for single
clusters. Hence already the first study comparing the two schemes in fits to 
LEP \ee data at 91.2 GeV obtained for the uncorrelated jet fit \CH~values twice 
as large as for the global scheme \cite{beca}. For our present study, the comparison 
is shown in table\ \ref{discrete} and indicates that \CH~is more than a factor of
two larger for the uncorrelated than for the global cluster scheme. Particularly,
the $\phi$ meson is found to deviate about $8\sigma$ from the data, which is
not the case
in the global cluster scheme (see table~\ref{fitee}); moreover, the temperature
value becomes much larger in the uncorrelated jet scheme. 
The reason for such
a behaviour is readily understood: in an uncorrelated jet scheme the phenomenon
of canonical suppression is enhanced by the requirement of strict local conservation
of quantum numbers and this drives the fit towards higher values of $T$ and/or 
$\gamma_S$ in order to reproduce the multiplicities; on the other hand, the $\phi$ 
meson is unaffected by canonical suppression and the increase of $T$ and/or 
$\gamma_S$ results in an overestimate of the production of this particle.

\begin{table}[!h]
\begin{center}
\begin{tabular}{|c|c|c|c|}
\hline
 Fit condition   & T (MeV) & $\gamma_S$ & $\chi^2$/dof   \\
\hline
Uncorrelated jet scheme           & 196.9 \tpm 1.74 &  0.622 \tpm 0.0096 &  104/12  \\
Global cluster scheme             & 164.8 \tpm 0.93 &  0.654 \tpm 0.0095 &  44/12  \\
\hline
\end{tabular}
\caption{Comparison of parameters for fits of 15 long-lived particles 
in \ee collisions at $\sqrt s = 91.2$ GeV, using different conservation scheme
for discrete quantum numbers. The errors within brackets are the fit errors rescaled 
by $\sqrt{\chi^2/dof}$ (see \cite{bcms}).}
\label{discrete}
\end{center}
\end{table}

\medskip

The mentioned quantum number distribution among the produced jets is only one
of the features to be specified for a concrete statistical model analysis
code. In addition, different codes involve further technical input details,  
addressing e.g. the little-known decay of heavy resonances, the mass range
of meson vs.\ baryon resonances to be included, etc. As a result, there exist
several codes, differing in these rather technical details. Since one cannot
give general physics grounds to judge one as ``better'' than another, the only
tool we have is to compare the \CH~they provide for specific data sets. For
the twelve species ``standard'' \ee data set at 91.2 GeV defined above in
Table III, we had obtained a \CH~=~41.5/9 $\simeq 4.6$. Using the same code,
we found \cite{bcms} for a much larger set of 30 species, including short-lived
resonances, a \CH~= 215/27 $\simeq 8$. This shows that also the choice of the
data set affects the resulting \CH; including broad resonances, with all the
resulting experimental problems, significantly increases \CH.

\medskip

Using a different code, ref.\ \cite{pbm} obtained for essentially the same
large data set at 91.2 GeV a fit with \CH~= 499/28 $\simeq 17.8$, i.\ e., a value twice
that which we had found. We can only conclude that the code used in ref.\ \cite{pbm}
must invoke physics features not in accord with the data. One such feature
was already indicated: while we use global cluster scheme, ref.\ \cite{pbm} retains
only the first term of the expansion shown in Fig.\ \ref{jets}, the
uncorrelated jet scheme. To look at further details, we consider a comparison 
proposed in ref.\ \cite{pbm}. Fixing the temperature $T=158$ MeV, the
strangeness suppression $\gs=0.8$ and the 
volume $V=30$ fm$^3$, a set of species' rates is calculated using the code of
\cite{pbm}, using our code, and using the publicly available code THERMUS \cite{thermus}.   
The results are shown in table \ref{confro}. At first sight, the 
output yields look fairly similar, and in particular those of our code and
those of THERMUS indeed show satisfactory agreement, with rather small and fluctuating
differences. A second look, however, shows that ref.~\cite{pbm}
predicts for almost all species a larger multiplicity than the other two.
This could be due to the inclusion of a larger number of heavy resonances. 
However, the relative differences are not uniformly distributed; the 
outstanding deviations are proton and $\Lambda$, which in ref.\ \cite{pbm}
lead to yields which are 1.5 - 1.7 times larger than those obtained in
the other two codes. These two states are among the most accurately
measured particles at LEP (3.5 \% error for the $p$, 2 \% for the $\Lambda$),
and hence a sizeable difference in the predicted yields will have a large impact 
on the final fit. We believe that this illustrates once more our main point: 
it is not the absolute value of \CH~that matters, but rather the average
relative deviation of fit to data. And here we expect that different codes 
shall (or should) lead to the same conclusion.

\begin{table}[!h]\begin{center}
\begin{tabular}{|c|c|c|c|}
\hline
                   &  Our code  & THERMUS  & Code  of ref.~\cite{pbm} \\                               
\hline

$\pi^0$            & 9.163     & 8.29      &  11.08     \\                                    
$\pi^+$            & 7.763     & 7.14      &  9.27     \\                                     
$K^+$              & 0.9953    & 0.945     &  1.014    \\                                     
$K_S^0$            & 0.9706    & 0.920     &  0.976     \\
$\eta$             & 1.047     & 0.890     &  1.090     \\
$\rho^0$           & 1.084     & 1.044     &  1.12      \\                    
$K^{0*}$           & 0.3126    & 0.285     &  0.299    \\                                 
$p$                & 0.2840    & 0.334     &  0.487    \\
$\phi$             & 0.1274    & 0.132     &  0.131    \\                                 
$\Lambda$          & 0.1170    & 0.120     &  0.182    \\                                 
$\Sigma^{*+}$      & 0.01512   & 0.0158    &  0.0197    \\                                
$\Xi^-$            & 0.00862   & 0.0095    &  0.0101    \\                                
$\Xi^{*0}$         & 0.000342  & 0.00340   &  0.00384    \\
$\Omega$           & 0.000541  & 0.000585  &  0.000625    \\
\hline
\end{tabular}\end{center}
\caption{Comparison between the output of our code, of THERMUS \cite{thermus},
and of the code used in ref.~\cite{pbm}, for a chosen set of species and 
fixed parameter values.}
\label{confro}
\end{table}

\section{Conclusions}

We have carried out statistical analyses of large multihadron production
samples from \pp, \AuAu and $e^+e^-$ interactions, containing the same
number of species and, as far as possible, the same kind of species. We 
use RHIC data from STAR ($\sqrt s = 200$ GeV) for the first two cases, LEP 
data averaged over the four CERN experiments ($\sqrt s = 91.25$ GeV) for the last. 
The main results are summarized in table~\ref{summary}. The hadronization
temperatures are seen to agree extremely well, and the average deviation
between the fit abundances and the data values (theory minus experiment/experiment)
are around 10\% for all three configurations.

\begin{table}[!h]
\begin{center}
\begin{tabular}{|c|c|c|c|}
\hline

      Collision                                 & \pp             & \AuAu            &  \ee   \\

\hline
\hline
~~~

Temperature [MeV]                               & 169.8 \tpm 4.2 &  168.5 \tpm 4.0 &  164.7 \tpm 0.9  \\
\hline 
Average relative deviation data vs.\ fit [\%]   & 12.5 &  11.7   &  9.4  \\
\hline
Average relative error of data [\%]             & 18   &  10     &  5.7  \\
\hline
\CH                                             & 15.0/8 $\simeq$ 1.9 &  22.2/8 $\simeq$ 2.8 & 41.5/9 $\simeq$ 4.6 \\
\hline
\hline
\end{tabular}
\caption{Summary of the fit results for a subset of 12 long-lived particles in 
high energy \pp, \AuAu and \ee collisions.}
\label{summary}
\end{center}
\end{table}


On the other hand, the resulting \CH~values are approximately 2 for \pp, 3 
for \AuAu and above 4 for $e^+e^-$. We argue that this does not 
imply a corresponding hierarchy of agreement with a statistical description. 
Since the average deviations of the fitted abundances are essentially the same 
in the three cases, the observed differences in \CH~values are rather a
reflection of the relative errors in the three experiments, also shown in
table~\ref{summary}. To test this, we have rescaled the errors
in the comparison \pp vs.\ \AuAu and in \ee vs. Au-Au, and in
both cases, the resulting \CH~values then become comparable.
We thus conclude that the hadroproduction abundances from high energy
\pp, \AuAu and \ee interactions agree equally well, i.e, to about
10\%, with the best present statistical model parametrization. 

\section*{References}

\end{document}